\documentclass[a4paper,aps,pre,superscriptaddress,floatfix,longbibliography,notitlepage,twocolumn]{revtex4-2}
\usepackage{bbm}
\usepackage{bbold}
\usepackage[pdftex]{graphicx}
\usepackage{latexsym,amsmath,verbatim,amssymb,txfonts}
\usepackage{lipsum}
\usepackage{color}
\usepackage{rotating}
\usepackage{multirow}
\usepackage[english]{babel}
\usepackage{comment}
\usepackage{bm}
\usepackage{amsfonts}
\usepackage{braket}
\usepackage{blkarray, bigstrut}
\usepackage{subcaption}

\begin{document}
\title{Collective dynamics of higher-order Vicsek model emerging from local conformity interactions}

\author{Iv\'an Le\'on}
\email{leonmerinoi@unican.es}
\affiliation{Department of Applied Mathematics and Computer Science, Universidad de Cantabria, Santander 39005, Spain}

\author{Riccardo Muolo}
\affiliation{RIKEN Center for Interdisciplinary Theoretical and Mathematical Sciences (iTHEMS), Saitama 351-0198, Japan}
\affiliation{Department of Systems and Control Engineering, Institute of Science Tokyo (former Tokyo Tech), Tokyo 152-8552, Japan}

\author{Hiroya Nakao}
\affiliation{Department of Systems and Control Engineering, Institute of Science Tokyo (former Tokyo Tech), Tokyo 152-8552, Japan}
\affiliation{International Research Frontiers Initiative, Institute of Science Tokyo (former Tokyo Tech), Kanagawa 226-8501, Japan}

\author{Keisuke Taga}
\email{tagakeisuke@rs.tus.ac.jp}
\affiliation{Department of Physics and Astronomy, Tokyo University of Science, Chiba 278-8510, Japan}
\affiliation{Department of Systems and Control Engineering, Institute of Science Tokyo (former Tokyo Tech), Tokyo 152-8552, Japan}

\date{\today}

\begin{abstract}
The Vicsek model is the paradigmatic framework for collective motion in systems of self-propelled particles. In its continuous-time formulation and most of its extensions, alignment arises from pairwise interactions among neighboring particles. In this work, we consider a model in which each particle assigns weights to its neighbors according to their alignment with the local consensus. This simple mechanism naturally yields a Vicsek-type model with pairwise and higher-order (i.e., nonpairwise) interactions. We analyze its collective dynamics through numerical simulations and approximate theory in the absence and presence of noise. The higher-order interactions generate a novel bidirectionally ordered phase, in which particles self-organize into oppositely moving groups. Moreover, we show that, depending on the relative strength of the pairwise and three-body interactions, the order-disorder transition can be either continuous or abrupt.
\end{abstract}

\maketitle
 \section{Introduction}

 	In nature, we often observe remarkable self-organization phenomena in which complex spatiotemporal behaviors emerge from interactions among elementary units~\cite{anderson1972more}. One such behavior is swarming, where groups of active particles (also called self-propelled particles) exhibit complex patterns~\cite{vicsek2012collective}. These particles can be inanimate, such as chemically driven particles, as well as living organisms, such as birds. Over the past years, researchers in statistical physics and nonlinear dynamics have proposed models and mechanisms to account for swarming behavior, from bird flocking to bacterial %diffusion 
	chemotactic patterns~\cite{reynolds1987flocks,helbing1995social,Nishiguchi2017long,peru12,Bech16,Cavagna2014,Marche13,chaintron2022propagation}.
	
	These phenomena can be studied through the so-called dry, aligning, dilute, active matter (DADAM) models~\cite{chate20}. 
	%These are
	They describe systems of active particles whose velocities follow a given alignment rule, 
	%with negligible repulsion between units (dilute), and without explicitly modeling the surrounding fluid (dry). 
	in which repulsion between units is neglected (dilute), and fluid surrounding the particles are not considered (dry).  
	The two most studied DADAM models are the polar DADAM model, in which particles preferentially align in parallel along a single direction (head--tail asymmetric), and the nematic DADAM model (or self-propelled rods)~\cite{GPBC10,Bar20}, in which particles 
	%align equally along a direction and its opposite (head--tail symmetric).
	tend to align equally in parallel and antiparallel directions (head--tail symmetric).
	
	The celebrated Vicsek model~\cite{vicsek1995novel}{\color{black}, in addition to being} the simplest polar DADAM model{\color{black}, captures} the transition from disorder to flocking~\cite{vicsek1995novel,toner1995long-range,gregoire2004onset,chate2008collective,ginelli2016physics}, and it has become a paradigmatic example of nonequilibrium statistical physics, analogously to the Ising model in equilibrium statistical physics~\cite{brush1967history,gallavotti2011statistical} and the Kuramoto model in synchronization dynamics~\cite{kuramoto1975,acebron2005kuramoto}.
	
    In the continuous-time Vicsek model, as well as in other DADAM models, the interactions are pairwise, i.e., the alignment is determined by a linear combination of two-body interactions. However, recent evidence across various fields, such as ecology and neuroscience, points to the importance of higher-order (i.e., nonpairwise) interactions in triggering and shaping collective behaviors~\cite{battiston2020networks,bianconi2021higher,natphys,bick2023higher,boccaletti2023structure,muolo2024turing,millan2025topology,battiston2026collective}. In particular, data-driven studies of animal groups suggest that collective motion may involve three-body effects or perception-mediated interaction rules~\cite{katz11,wang22,ITO24,shemesh13}, and in a wide variety of other systems~\cite{tanaka2011multistable,krawiecki2014chaotic,gambuzza2021stability,gallo2022synchronization,muolo2023turing,bick_explosive,skardal2020higher,millan2020explosive,carletti2020dynamical,iacopini2019simplicial,carletti2020random,della2023emergence,muolo2025higher,leon22a,dai2025higher}. Nevertheless, in active matter, only the swarmalator model has been extended to this framework~\cite{anwar2024collective,anwar2025two,hu2025effect}. The mechanism by which such interactions emerge is not yet fully understood, though, under certain conditions, they can be derived from pairwise interactions~\cite{leon19,bick2024higher,TAD2024} or delay~\cite{fujii2026emergence}.
	
	In this paper, we show that a continuous-time Vicsek model with higher-order (three-body) interactions naturally emerges from a DADAM model with local conformity interactions, in which the interaction weights depend on the degree of local alignment.
	The three-body interactions are not an \textit{ad hoc} addition to the alignment rule, but the consequence of the local conformity interactions across the particles.

	Our analysis indicates that higher-order interactions yield fundamental changes in the dynamics, including new stationary solutions and qualitatively different order--disorder transitions. In particular, we observe an abrupt change in the order parameter in comparison to the pairwise continuous-time Vicsek model, pointing to a distinct transition mechanism, that may be indicative of different critical behaviors and, possibly, a different universality class.
        
	The structure of the paper is as follows. In Section \ref{sec.DADAM}, we introduce the family of continuous-time weighted DADAM models. We show that the well-studied polar (Vicsek) and nematic DADAM models, and most of their extensions, belong to this family. In Section~\ref{sec.HOVic}, we demonstrate that by considering local conformity interactions weights, a Vicsek model with higher-order interactions is obtained. Sections \ref{sec.determin} and \ref{sec.noise} are devoted to the analysis of the dynamics of this model in the absence and presence of noise, respectively. Finally, in Section \ref{sec.conclusion}, we include the conclusions and discuss the implications of our results.

	\section{Weighted DADAM model}\label{sec.DADAM}
    In this Section, we first present the continuous-time weighted DADAM model and show that it contains the continuous-time Vicsek and nematic DADAM models as particular cases. We consider a system of $N$ active particles moving on a two-dimensional square of side $L$ with periodic boundary conditions, whose dynamics are given by
	\begin{eqnarray}
		\dot{\boldsymbol{x}}_j&=&\boldsymbol{v}_j,\label{eq.vicpos}\\
		\dot{\theta}_j&=&\frac{1}{n_j}\sum_{k\sim j}^N w_{jk}\sin(\theta_k-\theta_j)+\eta \xi_j(t),\label{eq.vicangle}
	\end{eqnarray}
	where $\boldsymbol{x}_j$, $\boldsymbol{v}_j$, and $\theta_j$
	%and $\boldsymbol{v}_j=v_0(\cos\theta_j,\sin\theta_j)$
	are the position, velocity, and direction of the particle $j$ ($j=1, ..., N)$, respectively.
	Equation~\eqref{eq.vicpos} describes the movement of particle $j$, where the velocity $\boldsymbol{v}_j=v_0(\cos\theta_j,\sin\theta_j)$ is determined by its direction $\theta_j$ and constant speed $v_0 \geq 0$.
	Equation~\eqref{eq.vicangle} describes the direction of particle $j$,
	%the alignment rule, 
	where the first term represents the alignment rule between particles $j$ and $k$, and the second term  represents independent Gaussian white noise. 
	%The weight $w_{jk}$ measures the strength with which particle $j$ follows particle $k$; larger $w_{jk}$ implies a stronger tendency of $j$ to match $k$. 
	The interactions among the particles are local, i.e., only particles within a radius $r_0$ affect each other. Thus, 
	%at any time,
	at any time,
	particle $j$ is influenced only by its $n_j$ neighbors satisfying $|\boldsymbol{x}_j-\boldsymbol{x}_k|\le r_0$. The sum is therefore performed only over such neighbors of $j$, as indicated by $k\sim j$\footnote{We restrict for simplicity the models to local and metric interactions, but metric-free interactions can be taken into account by simply summing over the appropriate neighbors.}.

	The weight $w_{jk}$, which {\color{black}may} depend on the positions and directions of all particles, determines 
	how strongly particle $j$ aligns with particle $k$; larger $w_{jk}$ implies a stronger tendency of particle $j$ to match the direction of particle $k$. Although we call this quantity weight due to its physical interpretation, they can be negative yielding antiferromagnetic interactions. As shown below, different choices of weights give rise to different families of DADAM models.
	The Gaussian white noise $\xi_j$ is characterized by $\langle\xi_j(t)\rangle=0$ and $\langle\xi_j(t)\xi_k(t')\rangle=\delta_{jk}\delta(t-t')$, while $\eta$ is the noise intensity. 

	The number of free parameters can be further reduced. Without loss of generality, we set $r_0=v_0=1$ by rescaling time and space. Since we are interested in bulk properties for large system sizes, the free parameters are the density $\rho=N/L^2$, the noise strength $\eta$, and the weights $w_{jk}$.
	
	\subsection{Continuous-time Vicsek model}
	
	The continuous-time Vicsek model~\cite{chate20} is a particular case of the weighted DADAM model, in which all weights are constant, i.e., $w_{jk}=\alpha \geq 0$:
	\begin{eqnarray}
		\dot{\theta}_j=\frac{\alpha}{n_j}\sum_{k\sim j}\sin(\theta_k-\theta_j)+\eta \xi_j(t)
		=\alpha r_j\sin(\psi_j-\theta_j)+\eta \xi_j(t),\label{eq.vicpairwise}
	\end{eqnarray}
	where, after the second equal sign, we defined the local alignment field
	\begin{equation}
		r_je^{i\psi_j}=\frac{1}{n_j}\sum_{k\sim j}e^{i\theta_k},
	\end{equation}
	where
	$0 \leq r_j \leq 1$ characterizes the local degree of alignment and $\psi_j$ represents its direction.
	It is then clear that, for $\alpha>0$, the interactions tend to align $\theta_j$ with the direction of the local field, $\psi_j$. 
	
	The aligning effect of the interaction is opposed by the diffusive effect of the noise term, so depending on which term dominates, the system exhibits either an ordered or disordered phase. Such phases can be characterized by the global polar order parameter
	\begin{equation}
		R=\frac{1}{v_0 N}\left|\sum_{k=1}^N \boldsymbol{v}_k\right|
		= \frac{1}{N}\left|\sum_{k=1}^N e^{i\theta_k}\right|.
	\end{equation}
	%This parameter gives the modulus of the center-of-mass velocity (polarization).
	This parameter measures the speed of the center-of-mass normalized by $v_0$, which quantifies the global polarization of the particles.
	For fixed density and strong noise, a disordered phase with $R\simeq 0$ is observed\footnote{Finite-size effects cause $R$ to fluctuate around zero.}. In 
	%that
	the disordered phase, the particle directions are approximately uniformly distributed because noise 
	dominates over the interactions.
	For weak noise, the interaction term dominates, yielding an ordered phase with $R>0$~\cite{chate20,vicsek1995novel,gregoire2004onset,chate2008collective,ginelli2016physics}. The transition between these two phases occurs through a liquid-gas transition in which some particles remain aligned in bands while other particles are randomly oriented \cite{chate20,sol15}. This transition is not continuous, though a very large number of particles and small densities are needed to perceive the discontinuous effect~\cite{gregoire2004onset,chate2008}.

	\subsection{Continuous-time nematic DADAM model}
	Another commonly studied model is the nematic DADAM model. Unlike to the Vicsek model, in which the particles have a preferential alignment direction, in the nematic model particles align equally in opposite directions. The model is recovered from Eq.~\eqref{eq.vicangle} if we adopt the weights of the form 
	\begin{equation}\label{eq.weightnematic}
		w_{jk}=2\alpha\cos(\theta_k-\theta_j),
	\end{equation}
 with $\alpha\geq 0$, yielding
	
	\begin{equation}
		\dot{\theta}_j=\frac{\alpha}{n_j}\sum_{k\sim j}\sin(2\theta_k-2\theta_j)+\eta \xi_j(t)=\alpha q_j\sin(\phi_j-2\theta_j)+\eta \xi_j(t).\label{eq.vicnematic}
	\end{equation}
	Here the local nematic field is defined as
	\begin{equation}
		q_je^{i\phi_j}=\frac{1}{n_j}\sum_{k\sim j}e^{2i\theta_k},
	\end{equation}
	where $0 \leq q_j \leq 1$ characterizes the local degree of {\color{black}nematic (bidirectional) alignment.}
	In this case, the interactions tend to align $\theta_j$ either with the direction of $\phi_j$ or $\phi_j+\pi$, i.e., the particles tend to align in parallel or in antiparallel with each other.
	Similarly to the Vicsek model, the nematic DADAM model exhibits a disordered phase for strong noise and a nematically ordered phase, in which particles move in two opposite directions, for weak noise~\cite{chate20,GPBC10,BMC08}. This transition is characterized by the global nematic order parameter
	\begin{equation}
		Q= \frac{1}{N}\left|\sum_{k=1}^N e^{i2\theta_k}\right|.
	\end{equation}
	The transition from the disordered state in the nematic DADAM model occurs through the emergence of bands of ordered particles. Those  bands become wider until they cover the whole space reaching the ordered phase~\cite{GPBC10,Bar20}.
	
	\subsection{Other models}

The weighted DADAM formulation also provides a convenient way to relate several extensions of Vicsek-type active-matter models. Different microscopic mechanisms can be incorporated by choosing suitable interaction weights \(w_{jk}\). For instance, mixtures of polar and nematic alignment can be obtained by summing the constant weight of Vicsek model with Eq.~\eqref{eq.weightnematic}, leading to competing ferromagnetic and nematic contributions~\cite{Ngo12}. Similarly, bounded-confidence interactions, in which particles interact with similarly oriented neighbors, correspond to weights that vanish or change sign when the angular difference between two particles exceeds a prescribed threshold~\cite{Romensky14}. 

Other modifications can also be represented within the same framework. Metric-free interactions can be incorporated by replacing the metric neighborhood \(k\sim j\) with a topological one~\cite{GC10,Pearce_2014}. Anisotropic, projected, or vision-based interactions~\cite{PMR14,BP16} can be modeled by allowing $w_{jk}$ to depend on the relative position of particle $k$ with respect to the heading or visual cone of particle $j$. Moreover, vectorial or temporally correlated noise can be interpreted as introducing stochastic or time-dependent contributions to the effective weights~\cite{gregoire2004onset,GBR18}, while delayed interactions can be incorporated by allowing the alignment angles on the weights to depend on past particle configurations~\cite{HH25}.

Thus, the weighted DADAM model should be understood as a broad class of continuous-time active-matter models in which different physical, perceptual, or social mechanisms are encoded through the choice of $w_{jk}$. In the next section, we exploit this flexibility to introduce local conformity interactions giving rise to higher-order interactions. 
    
	\section{Higher-order Vicsek model} \label{sec.HOVic}
	
	The continuous-time Vicsek model, the nematic model and others presented in the previous section have the particularity that the interactions (and weights) only depend on the alignment angles of two particles, i.e., they are pairwise. In this Section, we show that, from physically grounded weights, we can derive a Vicsek model whose interactions depend on three angles, that is, a Vicsek model with higher-order (i.e., nonpairwise) interactions.
	
\begin{figure*}[t]
		\centering
		\includegraphics[width=0.8\linewidth]{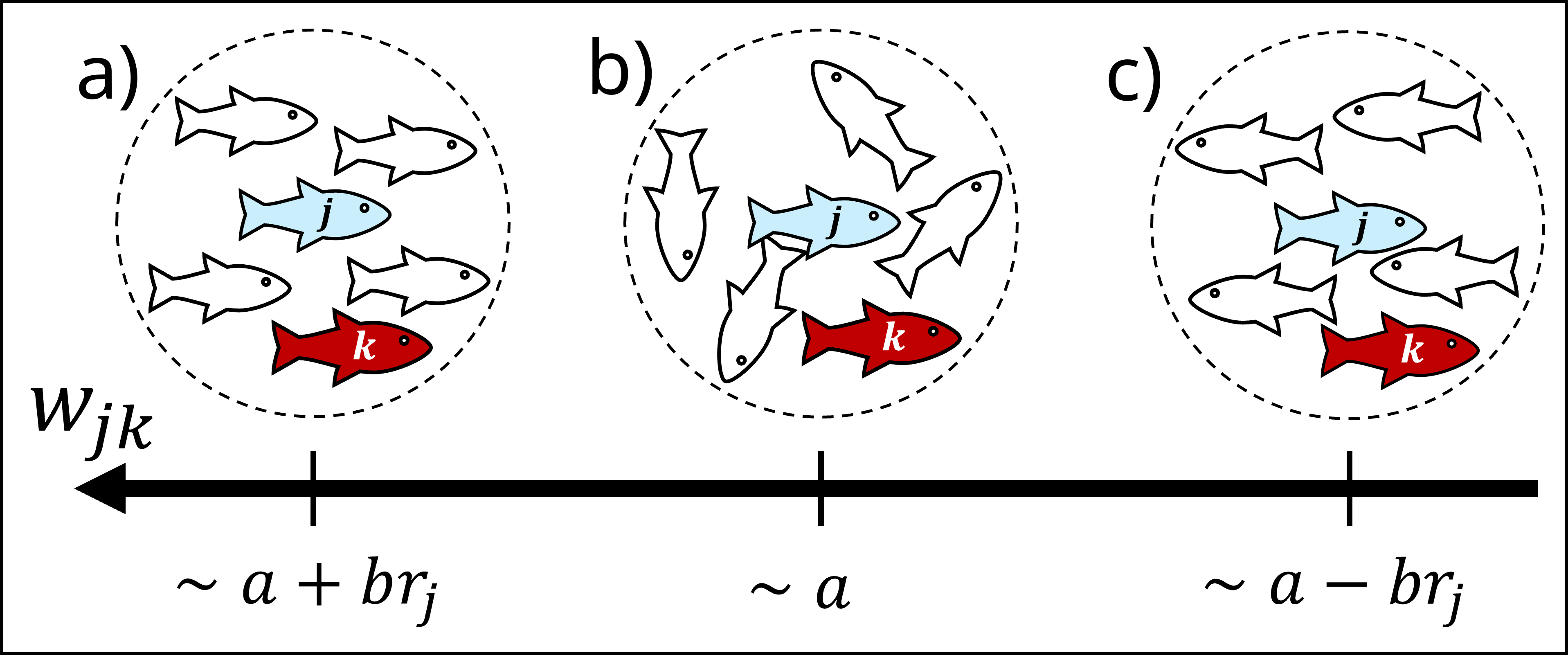}
		\caption{
			Schematic illustration of the local conformity weight $w_{jk}$ (for positive $a$ and $b$), which quantifies the influence of particle $k$ on particle $j$ based on the local alignment of particles around $j$. (a) High $w_{jk}$: particle $k$ is aligned with the local majority. (b) Intermediate $w_{jk}$: no dominant orientation exists locally. (c) Low $w_{jk}$: particle $k$ is oriented opposite to the local majority.}
		\label{fig.weight}
	\end{figure*}

	We consider local conformity interactions among particles, i.e., an interaction in which each particle tends to align to the neighbors that are more closely aligned to the local majority. These conformity interactions are described by the interaction weights
	\begin{eqnarray}\label{eq.weight}
		w_{jk}=a+\frac{b}{n_j}\sum_{l\sim j}\cos(\theta_l-\theta_k)=a+b r_j \cos(\psi_j-\theta_k),
	\end{eqnarray}
	where $a$ and $b$ are constants, which can be both positive and negative.
	The first term corresponds to the weight of the classical Vicsek interaction as in Eq.~\eqref{eq.vicpairwise},
	while the second term quantifies the alignment of particle $k$ with the direction of the local field $\psi_j$. 
	% (unlike the nematic weight)
	For $b>0$, particle $j$ assigns larger weights to neighbors whose movement direction is closer to the direction of the local majority, as schematically shown in Fig.~\ref{fig.weight}. 
	%This weight models active particles that do not follow all neighbors equally, 
	This describes a conformity interaction, since each active particle does not follow all the neighbors equally, but, instead, preferentially follows those closer to the local majority. For $b<0$ the argument is reversed and the particles avoid the local majority alignment. It is useful to note that the weights take values in the interval $w_{jk}\in[a-b,a+b]$. Depending on the relative values of $a$ and $b$, the interaction may be purely ferromagnetic, antiferromagnetic, or a combination of both.
	
	We remark that the second term in the weight Eq.~\eqref{eq.weight} is different from the nematic weight defined by Eq.~\eqref{eq.weightnematic}. Precisely this second term encodes the higher-order (namely, three-body) interaction, i.e., the weight that $j$ assigns to $k$ depends on the alignment angle to particle $l$.
	
	Substituting Eq.~\eqref{eq.weight} into Eq.~\eqref{eq.vicangle} yields
	\begin{eqnarray}
		\dot{\theta}_j&=&\frac{\alpha }{n_j}\sum_{k\sim j} \sin(\theta_k-\theta_j)
		+ \frac{\beta}{n_j^2}\sum_{k,l\sim j} \sin(2\theta_k-\theta_l-\theta_j)+\eta\xi_j(t)\nonumber\\
		&=&\alpha r_j \sin(\psi_j-\theta_j)+ \beta r_j q_j \sin(\phi_j-\psi_j-\theta_j)+\eta\xi_j(t),\label{eq.angthreebody}
	\end{eqnarray}
	where $\alpha=a+\frac{b}{2}$ and $\beta=\frac{b}{2}$.
	
	%Equation~\eqref{eq.angthreebody} 
	This equation contains a pairwise term representing the standard Vicsek alignment and a higher-order interaction term arising from the local conformity rule.
	These terms represent confronting effects: the first tends to align $\theta_j$ with $\psi_j$, while the second tends to align $\theta_j$ with $\phi_j-\psi_j$. Notice the difference of this last term in comparison to the nematic interaction terms that align the angle $\theta_j$ either to $\phi_j$ or $\phi_j+\pi$. Thus, the model is different from a Vicsek model with both polar and nematic interaction as in \cite{Ngo12,Hub18}, and also different to a combination of ferromagnetic and antiferromagnetic interactions~\cite{Chat23}. The signature of the higher-order interaction is the appearance of the product of the local fields $r$ and $q$, that cannot emerge by means of the sole pairwise interactions. We remark that, for $\beta=0$, the classical pairwise Vicsek model~\eqref{eq.vicpairwise} is recovered, while a Vicsek model with only higher-order interactions is obtained for $\alpha=0$.

	Throughout this paper, it can be useful to relate the parameters $\alpha$ and $\beta$ to the weights $w_{jk}$, c.f., Eq.~\eqref{eq.weight}, especially noticing that they take values on a finite range given by	
	\begin{equation}\label{eq.weightalphabeta}
		w_{jk}\in[\alpha-3\beta,\alpha+\beta].
	\end{equation}
	Hence, depending on the values of $\alpha$ and $\beta$, the interactions will be purely ferromagnetic/antiferromagnetic, or a combination of both.

	It is important to note that the higher-order interactions in Eq.~\eqref{eq.angthreebody} are not introduced as an \textit{ad hoc} higher-order alignment rule. Instead, they emerge naturally from the social-inspired conformity-based weight in Eq.~\eqref{eq.weight}. 
	
	In the context of the higher-order interactions, it is worth clarifying a few points in the difference between pairwise and higher-order (i.e., nonpairwise) interactions. The classical Vicsek model~\cite{vicsek1995novel} does involve nonpairwise interactions, as it is formulated in discrete time and the alignment is updated according to the average direction, which is a nonlinear function of all neighboring angles (see~\cite{Ihle11} for further discussion). By contrast, in the continuous-time formulation, the alignment arises from interactions depending only on two angle differences. In our considered model, Eq.~\eqref{eq.angthreebody}, the interaction depends nonlinearly on three different angles, i.e., they are higher-order. One can notice that the normalization by the number of neighbors $n_j$ introduces a dependence on the full neighborhood in the interaction. However, this dependence acts as a common pre-factor for all interactions of the particle $j$, and it is, therefore, not expected to play a significant role. In particular, such a factor could be eliminated by considering metric-free interactions~\cite{GC10,Pearce_2014}. 

	In what follows, we study the collective dynamics of the higher-order Vicsek model,  Eqs.~\eqref{eq.vicpos} and \eqref{eq.angthreebody}. Let us note that there are four control parameters: the density, that we fixed to $\rho=2$, the noise intensity $\eta$, and the two- and three-body coupling strengths $\alpha$ and $\beta$.

	\section{Deterministic case} \label{sec.determin}
    
	We begin our study analyzing the deterministic case, i.e., $\eta=0$. In the absence of noise, the relative strength of $\alpha$ and $\beta$ completely determines the asymptotic behavior. We first perform numerical simulations to study the dynamics. Then, we derive a theory for the global interaction case, i.e., $r_0\to\infty$ (equivalent to a mean-field approach). As shown below, such an approximate theory captures qualitatively the dynamics of the higher-order Vicsek model.
	
	\subsection{Numerical results}
\begin{figure*}[t]
	\centering
		\includegraphics[width=0.95\linewidth]{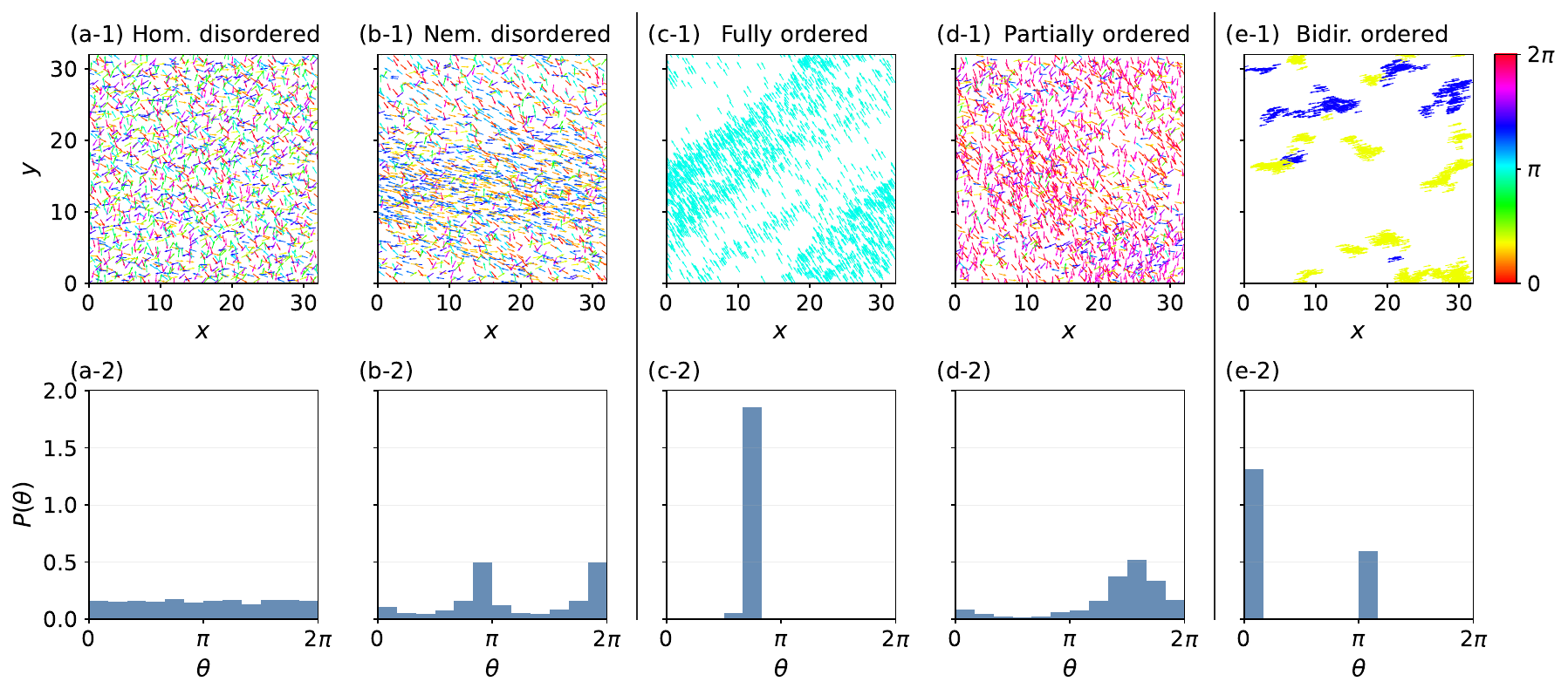}

	\caption{
		Top row: snapshots of the deterministic dynamics ($\eta=0$) for system size $L=32$ (videos can be found in the SM). Bottom row: probability distribution of angles. Each column represents a different phase: 
		(a) homogeneous disordered phase at $\alpha=-1$, $\beta=1$;
		(b) nematic disordered phase at $\alpha=-1$, $\beta=-1$;
		(c) fully ordered phase at $\alpha=1$, $\beta=-1$;
		(d) partially ordered phase at $\alpha=0.5$, $\beta=-1$;
        (e) bidirectionally ordered phase at $\alpha=0$, $\beta=1$.
	}
	\label{fig.snapshot}
\end{figure*}

	 In our numerical simulations we use an Euler algorithm (or Euler-Maruyama algorithm in the stochastic case) with a time step of $\Delta t=10^{-2}$ for $N=2048$ particles in a square box of linear size $L=32$. Three general distinct phases can be observed in the simulations: disordered, ordered, and bidirectionally ordered, as shown in Fig.~\ref{fig.snapshot}.

	The disordered phase is characterized by  a vanishing polar order parameter, i.e., $R\simeq 0$. Two qualitatively different cases of disordered phase can be identified.
	The first case is the homogeneous disordered phase, where the particle directions are approximately uniformly distributed, as shown in Fig.~\ref{fig.snapshot}(a-1) and (a-2).
	In this case, the local alignment, the local nematic field, and the global nematic order parameter vanish, i.e., $r_j \simeq q_j\simeq Q\simeq 0$. 
	%Nevertheless, 
	The second disordered phase is the nematic-disordered phase, in which the distribution of the particle directions is non-uniform, as shown in Fig.~\ref{fig.snapshot}(b-1) and (b-2). We refer to such phase as disordered since $r_j\simeq0$ and $R\simeq 0$, yet the alignment is non-uniformly distributed, as indicated by the non-zero nematic order parameter, $0 < Q < 1$, shown by the distributions of angles in panel (b-2). Both homogeneous and non-uniform disordered phases are achieved because the interaction in Eq.~\eqref{eq.angthreebody} is proportional to $r_j$, which vanishes in both disordered phases. Our numerical analysis reveals that the nematic-disordered phase is stable only when the homogeneous disorder is also stable.
	
	The fully ordered phase is characterized by particles moving approximately in the same direction, as shown in Fig.~\ref{fig.snapshot}(c-1) and (c-2),
	meaning that the local fields and order parameters are close to maximal, i.e., $R\simeq Q\simeq r_j\simeq q_j\simeq 1$. 
	We 
	%highlight
	note that, for some parameter values, the competition between pairwise and three-body interactions can also yield a partially ordered phase, even in the absence of noise, Fig.~\ref{fig.snapshot}(d-1) and (d-2). 
	In this phase, both the polar and nematic order parameters are finite, i.e., $0 < R< 1$ and $0< Q < 1$. We remark that, unlike to the Vicsek model where disruption of complete order is due to the presence of noise, here partial order emerges from the competing effect of pairwise and higher-order interactions.

	While ordered and disordered phases are also observed in the classical Vicsek model, three-body interactions give rise to a distinct third stationary behavior: bidirectional ordered phase. This phase is characterized by some of the particles moving in one direction and the others moving in the opposite direction, as in Fig.~\ref{fig.snapshot}(e-1) and (e-2), yielding $ R<1$ while $Q\simeq 1$. In the pairwise Vicsek model, Eq.~\eqref{eq.vicpairwise}, bidirectional order is an unstable solution. The presence of the higher-order interactions stabilizes this phase and makes it reachable from random initial conditions. This phase is similar to the apolar-ordered phase observed in the Vicsek model with bounded confidence~\cite{Romensky14}, though in that model small clusters moving in perpendicular directions also appear. 

    	\begin{figure*}[t]
	\centering
		\includegraphics[width=0.95\linewidth]{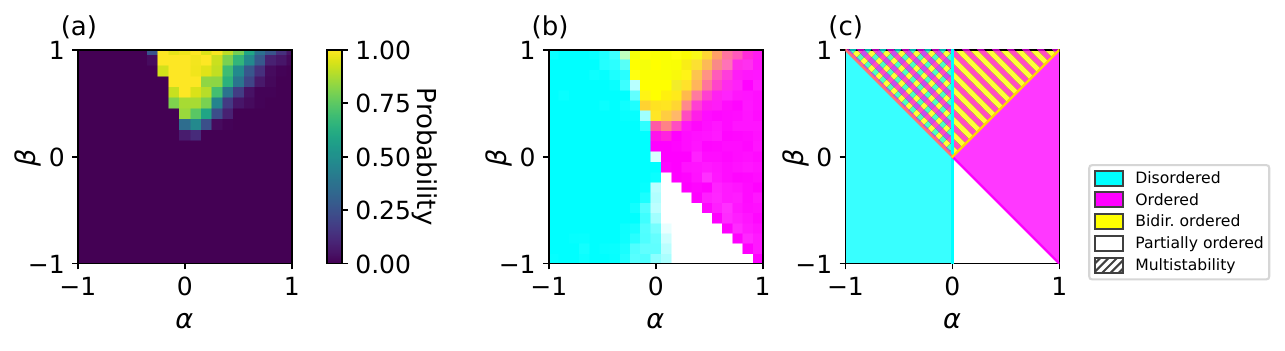}

	\caption{
		(a) Probability of reaching bidirectional order from random initial conditions.
		(b) Partial phase diagram of the asymptotic state reached from random initial conditions: blue (disorder), yellow (bidirectional order), pink (full order), white (partial order). The stationary state was classified by measuring the average alignment and nematic order parameter over $t=10$ t.u. after a transient time was discarded. The asymptotic phase was classified as fully-ordered phase if $R \ge 0.99$, as disordered phase if $R < 0.01$ and $Q < 0.99$, as bidirectional ordered phase if $R < 0.99$ and $Q > 0.99$, or as partially ordered phase otherwise.		
		(c) Theoretical phase diagram in the global-coupling limit $r_0\to\infty$ (same color code); hatched regions indicate multistability.
	}
	\label{fig.noiseless}
\end{figure*}

	To quantify how frequently bidirectional order is realized, we depict in Fig.~\ref{fig.noiseless}~(a)
	the probability of reaching 
	bidirectional order for each pair $(\alpha,\beta)$. For each parameter pair, we simulated the system from $100$ random initial conditions and measured the fraction that converges to bidirectional order. We observe that, for $\beta>0$ and $\alpha$ near zero, bidirectional order is frequently 
	realized. Since these parameters correspond to strong higher-order interactions and weak (or absent) pairwise interactions, the observed bidirectional order  
	is clearly an effect of higher-order interactions.
	
	To provide a global view of the system behavior, in Fig.~\ref{fig.noiseless}~(b), we show the corresponding partial phase diagram obtained from $100$ random initial conditions. For each pair of values $(\alpha,\beta)$, we depict in blue, pink, yellow, or white the stationary state reached: disordered, fully ordered, bidirectional ordered, or partially ordered, respectively. At parameter values for which different initial conditions lead to distinct stationary states, the colors are combined proportionally to the frequency with which each state is observed.    
	Note that this is not a complete phase diagram, because we only consider random initial conditions; in particular, regions of bistability between the disordered and ordered phases are not captured. Such bistability regions can be observed in the numerical simulations by considering nearly aligned initial conditions. We remark that, in the region where the interaction is purely ferromagnetic, i.e., $\alpha-3\beta>0$ and $\alpha+\beta>0$ c.f. Eq.~\eqref{eq.weightalphabeta}, only the fully ordered phase is stable. Conversely, in the region where the interaction is purely antiferromagnetic, only the disordered phase is stable. Therefore, the coexistence of both types of interactions is responsible for the emergence of a more complex landscape. 
	
	\subsection{Global interaction theory}
	Additional insight can be gained by analyzing the global-coupling or mean-field limit
	$r_0\to\infty$, where particle positions become irrelevant and the dynamics depend only on the directions. The analysis of this model is equivalent to a mean-field analysis \cite{Grob13,Peruani2008}, thus it is able to capture the phases in which particles are spatially uniformly distributed. 
	
	When the interactions are considered global, the position of the particles are no longer relevant and the dynamics reduce to an extension of the Kuramoto model with higher order interactions~\cite{kuramoto1975,Kuramoto_book,acebron2005kuramoto}. Using standard techniques, analytical results in the $N\to\infty$ limit can be derived, and are summarized in the phase diagram of Fig.~\ref{fig.noiseless}(c). The full analysis is provided in the Appendix \ref{app.detkur} and in this subsection we solely comment the main results. 
    
    According to the analysis, the disordered phase is stable for $\alpha<0$ (blue region), while the fully ordered phase is stable when $\alpha+\beta>0$ (pink region). The bidirectional ordered phase is not properly captured by the global-coupling model because this phase is not spatially uniform. However, by taking into account the presence of spatial clusters we obtain that the stability depends on two exponents $\lambda_1\propto-(\alpha+\beta)$ and $\lambda_2\propto\alpha-\beta$. Thus bidirectional order is stable when $\alpha+\beta\ge 0$ and $\alpha-\beta\le 0$ in the presence of spatial inhomogeneities (yellow region). Finally, the theory predicts the presence of a partially ordered phase at $\alpha+\beta<0$ with $\alpha>0$ (white region),  similar to self-consistent partial synchrony (or quasiperiodic partial synchrony) observed in populations of oscillators~\cite{CPR16,CP19,leon2020}. The phase diagram reveals regions of multistability between different phases, depicted using hatching. Such multistability was also  observed in the numerical simulations of the local interaction model, though it is not appreciable in Fig.~\ref{fig.noiseless}(b) since only random initial conditions were considered.
	
	The analysis of the global-coupling limit also provides 
	insights into the probability of reaching the bidirectional ordered phase: regions where bidirectional order is more likely coincide with parameter values for which the sum of the stability exponents is more negative, 
	suggesting a larger basin of attraction.
	
	Comparison of Fig.~\ref{fig.noiseless}(b) and Fig.~\ref{fig.noiseless}(c) indicates that the global-coupling limit captures key qualitative features and serves as a useful first approximation, even though it does not fully describe the local-interaction model.

	\section{Effect of noise} \label{sec.noise}
	
	In this section we study how the inclusion of noise affects the dynamics. In particular, we are interested in the transition between the ordered and disordered phases. Analogously to the previous section, we first perform numerical simulations and then develop the theory for the global interaction limit.  
	
	\subsection{Numerical results}
		\begin{figure*}[t]
		\centering
		\includegraphics[width=2\columnwidth]{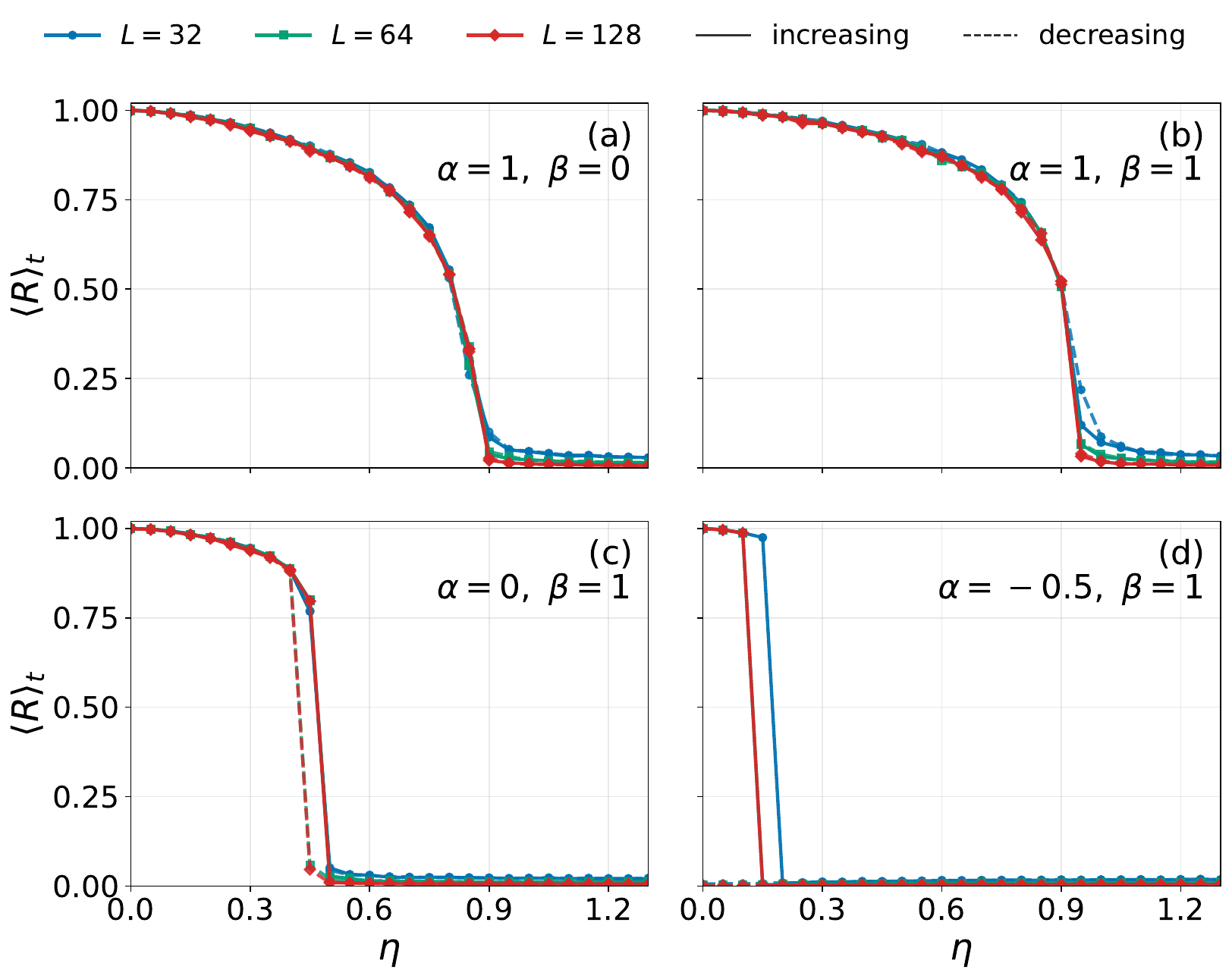}\\
		\caption{Global polar order parameter $\langle R\rangle_t$ averaged over time for different pairs $(\alpha,\beta)$ as the noise intensity $\eta$ is increased (solid line) or decreased (dashed line), for different system sizes. Blue, green and red lines correspond to $L=32$, $64$ and $128$ respectuvely, corresponding to system sizes of $N=2048$, $8192$ and $32768$ particles, respectively.}
		\label{fig.bifdiag}
	\end{figure*}
	
	The noise in the higher-order Vicsek model adds diffusion to the alignment angles of the particles, affecting the character and stability of the phases described in the previous section. Using numerical simulations, we study the effect of noise on the disordered, ordered, and bidirectionally ordered phases observed in the noiseless case.
	
	Since the noise adds some diffusion to the alignment directions, it preserves the disordered phase. Notice however, that noise tends to flatten the distribution of alignment angles, so nematic disorder takes the minimum possible value of the nematic order parameter $Q$. Since nematic disorder was stable for parameters in which homogeneous disorder was already stable, the noise completely homogenizes the directions reaching the homogeneous disordered phase. Thus, nematic disorder is not achieved for any amount of noise.  
		
	The bidirectional ordered phase is only observed in our simulations for very weak noise. A careful numerical study shows that this phase persists for weak noise, and vanishes when noise is increased. This is similar to the apolar ordered phase observed in~\cite{Romensky14} that is only displayed for weak noise.
	 
	The introduction of noise strongly affects the fully ordered phase by preventing perfect alignment of the particles.
	Thus, the polar order parameter $R$ 
	no longer takes the maximal value $1$,
	and, for sufficiently strong noise, the ordered phase disappears, as evidenced by $R$ reaching zero.
	
	\begin{figure*}[t]
		\centering
		\includegraphics[width=2\columnwidth]{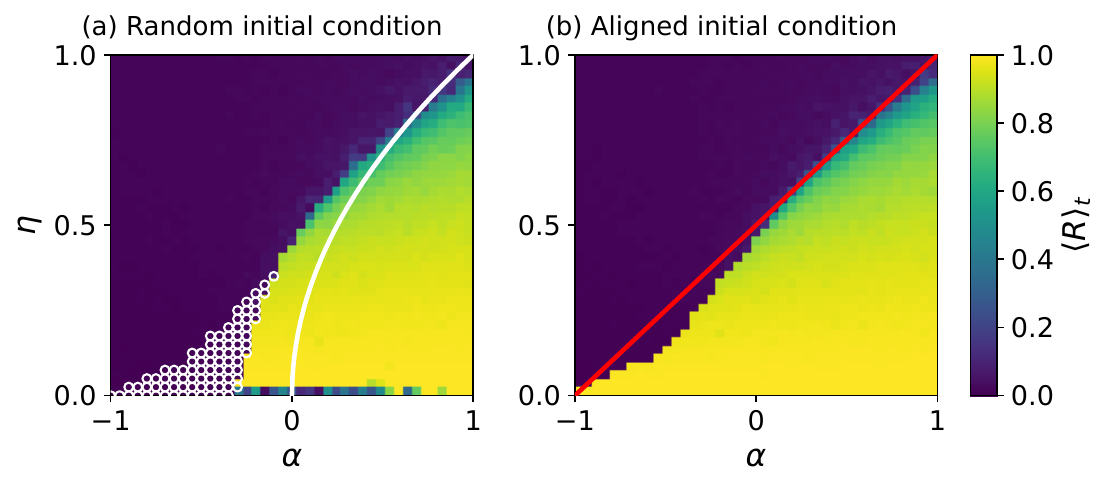}\\
		\caption{Phase diagram of the higher-order Vicsek model,
			% with higher-order interactions,
			Eqs.~\eqref{eq.vicpos} and \eqref{eq.angthreebody}, for fixed $\beta=1$ and $L=32$. The color scale shows the time-averaged polar order parameter $\langle R\rangle_t$ when the system is initialized with random angles (a) or nearly identical angles (b). White circles are drawn to indicate region of bistability between ordered and disordered phases. White and red lines mark the theoretical stability boundaries of disordered and ordered phase, respectively, in the global-coupling limit.}
		\label{fig.phasediag}
	\end{figure*}
	To characterize the order--disorder transition, we compute the time-averaged polar order parameter $\langle R\rangle_t$ over $5\cdot10^4$ t.u., while varying the noise intensity $\eta$ for fixed $\alpha$ and $\beta$. We consider four representative choices: the pairwise case $(\alpha,\beta)=(1,0)$, the pure three-body case $(\alpha,\beta)=(0,1)$, and two mixed cases, namely, $(\alpha,\beta)=(-0.5,1)$ and $(\alpha,\beta)=(1,1)$. For each pair, we start from $\eta=0$ ($\eta=1.3$) with completely aligned initial conditions and adiabatically increase (decrease) $\eta$ while measuring $\langle R\rangle_t$ after a long enough transient. The results are shown in Fig.~\ref{fig.bifdiag}.

    The numerical results indicate that, when only pairwise interactions are present (Fig.~\ref{fig.bifdiag}(a)) or when they are strong (Fig.~\ref{fig.bifdiag}b), the order parameter departs smoothly from zero. This means that, when pairwise interactions are dominant, the system behaves similarly to the classical Vicsek model\footnote{As in the pairwise Vicsek model, we expect a discontinuous transition in the $N\to\infty$ limit, with an inappreciable jump in $\langle R\rangle_t$ for the considered number of particles~\cite{gregoire2004onset,chate2008}.}. 
    
	By contrast, when three-body interactions dominate, Figs.~\ref{fig.bifdiag} (c--d), the competition between pairwise and higher-order terms yields an abrupt change in $\langle R\rangle_t$. This means that the system switches abruptly from a completely incoherent phase to a phase in which a large number of units are flocking. Furthermore, in Fig.~\ref{fig.bifdiag}(d), we also observe a region of bistability between order and disorder. There, an ordered or disordered phase is achieved depending on the initial conditions. 
    
    Finally, for sufficiently negative $\alpha$, the disordered phase is already stable in the absence of noise (see Fig.~\ref{fig.noiseless}(b)), so no transition occurs as $\eta$ is increased.
	
	We can further explore how pairwise and higher-order interactions shape the collective dynamics through the partial phase diagrams, depicted in Fig.~\ref{fig.phasediag}. These diagrams show the value $\langle R\rangle_t$, in color code, for different pairs $(\alpha,\eta)$ at fixed $\beta=1$. At each point, we run simulations initialized either with random directions Fig.~\ref{fig.phasediag}~(a) or nearly identical directions Fig.~\ref{fig.phasediag}~(b), and measure $\langle R\rangle_t$ in the steady state after a long enough transient is discarded. In panel (a), white circles are depicted to better visualize the numerically found bistable region between the ordered and disordered phases. 
	These phase diagrams show that positive $\alpha$ enhances order, which is expected because both terms in Eq.~\eqref{eq.angthreebody} promote alignment. For negative $\alpha$, pairwise and three-body interactions oppose each other, leading to bistable regions. One can observe that, in Fig.~\ref{fig.phasediag}~(a), for very weak noise, there are points in the parameter space for which the polar order parameter $R$ reaches an almost vanishing value, that gets close to $1$ for slightly stronger noise intensities. This behavior is not an artifact but rather a consequence of the deterministic dynamics: in that region, the system settles into the bidirectional order phase with a large basin of attraction (see Fig.~\ref{fig.noiseless}(a)) and very small polar order parameter.
	
	In terms of the coefficients of the interaction weights , $a$ and $b$, c.f. Eq.~\eqref{eq.weight}, we have a similar result as in the deterministic case: if the weights are purely ferromagnetic, we have a transition similar to the classic Vicsek model, while, for purely antiferromagnetic weights, the disordered phase is observed for any amount of noise. We note that the presence of both ferromagnetic and antiferromagnetic weights is not solely responsible for the abrupt transition. For example, in Fig.~\ref{fig.bifdiag}(b), we observe that the order-disorder transition is smooth even though the weights take values in the range $[-2,2]$. We also highlight that, for all points of the phase diagram of Fig.~\ref{fig.phasediag}, the system contains a mixture of ferromagnetic and antiferromagnetic weights. 
		
	\subsection{Global interaction theory}
    In this Section, we comment the results of the analysis of the global-interaction limit of higher-order Vicsek model. This theory, equivalent to a mean-field theory, provides a useful intuition on the dynamics and mechanism producing the stationary phases. As discussed before, since the interactions are global, the dynamics are reduced to the stochastic Kuramoto model with higher-order interactions. For this model, one can write the Fokker-Planck equation, find the stationary solutions and analyze their stability. In this section we discuss the main results and keep the analysis and details in the Appendix \ref{app.stochkur}.
	
	The only disordered solution to the stochastic Kuramoto model with higher-order interactions is the homogeneous disordered phase. The stability analysis indicates that the disordered phase loses stability at $\eta=\sqrt{\alpha}$, corresponding to the white line in Fig.~\ref{fig.phasediag}(a).
	
	The global-interaction limit model also captures the partially ordered solution, that becomes stable at
	\begin{equation}\label{eq.partordbound}
		\eta=\begin{cases}
			\dfrac{\alpha+\beta}{2\sqrt{\beta}}& \text{if } \alpha<\beta, \\
			\sqrt{\alpha} & \text{if } \alpha\ge\beta,
		\end{cases}
	\end{equation}
	that has been depicted in red in Fig.~\ref{fig.phasediag}(b). This solution can emerge through two distinct bifurcation mechanism. For $\alpha<\beta$, it arises via a saddle-node bifurcation, leading to an abrupt transition (as in \cite{skardal2019abrupt,leon2024,MK25}) corresponding to Fig.~\ref{fig.bifdiag}(c--d) of the higher-order Vicsek model. By contrast for $\alpha\ge\beta$, the disordered state becomes unstable through a pitchfork bifurcation, leading to a continuous transition, as occurs in Fig.~\ref{fig.bifdiag}(a--b) for the higher-order Vicsek model. The corresponding bifurcation diagrams for the global coupling limit can be found in Appendix \ref{app.stochkur}.	

	The numerical results in Figs.~\ref{fig.bifdiag} and \ref{fig.phasediag} reveal an order–disorder transition that differs quantitatively from those of polar (Vicsek) and nematic DADAM models. Furthermore, the global interaction or mean-field analysis indicates that the system behaves differently in the presence of higher-order interactions. Taken together, this evidence suggests that the higher-order Vicsek model, Eq.~\eqref{eq.angthreebody}, may exhibit distinct critical behavior compared to Vicsek or pairwise DADAM models. Whether this corresponds to a separate universality class or not remains an open question.
	
	\section{Discussion and conclusions} \label{sec.conclusion}

    In this work, we have studied a Vicsek-type model with both pairwise and higher-order (i.e., nonpairwise) interactions. This model was derived by considering a weighted DADAM model in which each particle weights its neighbors according to a conformity rule. Such conformity rule naturally leads to the emergence of higher-order interactions. We have numerically investigated the dynamics of the system and compared it with the analytical results of the global interaction (or mean-field) limit. 

    We first considered the deterministic case, which is completely governed by the alignment interaction. In this case, depending on the strength of the pairwise and higher-order interactions, we observed three different asymptotic phases: disorder, order, and bidirectional order. We highlighted the bidirectional ordered phase in which particles move in clusters of opposite directions induced by the higher-order interactions. This phase is similar to the two-cluster states observed in phase oscillator models and the higher-order Kuramoto model~\cite{Golomb92,HMM93,KK2001,tanaka2011multistable,leon2024}. We also remarked that, even in the absence of noise, the system achieves a partial order phase due to the competing effect of both interaction types.

    We then studied the effect of noise on the dynamics, with particular focus on the order-disorder transition. The numerical simulations and mean-field theory showed that the system presents a smooth transition from order to disorder for strong pairwise interactions, or an abrupt transition when higher-order interactions are dominant. Furthermore, regions of bistability between the disordered and ordered phases are observed. This suggests that the  higher-order Vicsek model exhibits a different critical behavior from the pairwise one. Let us remark that, even if in the $N\to\infty$ limit, the critical behavior was equivalent, real-world scenarios and experiments involve finite populations. Hence, the difference in the behavior induced by the presence of higher-order interactions may be important in describing real-world systems behaviors, leading to more accurate modeling descriptions.

	Our global-coupling theory provides a qualitative description, nevertheless, a complete theory may require on hydrodynamic approaches, as in~\cite{toner1995long-range,Bertin06,Bertin13} for polar (Vicsek) and nematic DADAM models. A better understanding of the critical behavior would also help clarify whether Eq.~\eqref{eq.angthreebody} belongs to a different universality class. Finally, the order-disorder transitions in DADAM models are known to occur through the emergence of bands~\cite{toner1995long-range,GPBC10}. Our simulations of the higher-order Vicsek model are consistent with this scenario, {\color{black}though}
	%But finite-size effects can be misleading in such problems, requiring further investigation.
	further investigation is required to clarify the finite-size effects on the transitions.

    Finally, our results provide a possible theoretical link with experimental observations of nonpairwise interactions in animal groups. Data-driven studies of fish schools have shown that individual responses may involve three-body effects or perception-mediated interaction rules \cite{katz11,wang22,ITO24}, while higher-order social interactions have also been reported in groups of mice \cite{shemesh13}. In this context, the local conformity mechanism studied here offers a minimal active-matter framework in which effective nonpairwise alignment emerges naturally from context-dependent local interactions.
	
	The present model can be extended in multiple ways, such as by considering metric-free interactions~\cite{GC10,Pearce_2014}, finite vision cones~\cite{PMR14,BP16,napoli2026nest}, or time delays~\cite{HH25}. Other directions to explore would be understanding the effect of spatial dimension~\cite{CVV99,chate2008collective} and considering vector or correlated noise~\cite{gregoire2004onset,GBR18}. Understanding how higher-order interactions affect Vicsek-like models is, in our view, a key step towards a more complete comprehension of swarming behavior.

	\section*{Acknowledgments}
	The authors are grateful to Diego Paz\'o and Juan Manuel L\'opez for fruitful discussions. I.L. acknowledges support by MICIU/AEI/10.13039/501100011033 through Grant No.~PID2021-125543NB-I00 and by ERDF/EU, and thanks the Institute of Science Tokyo for the hospitality. R.M. acknowledges JSPS KAKENHI 24KF0211. H.N. acknowledges JSPS KAKENHI 25H01468 and 25K03081. K.T. acknowledges JSPS KAKENHI 24K20863 and Universidad de Cantabria for the hospitality. We acknowledge the 2025 International Active Matter conference held at Meiji University (Japan) in Jan 2025 (supported by MEXT Promotion of Distinctive Joint Usage/Research Center Support Program Grant Number JPMXP0724020292), where the idea of this work originated.

    \section*{Author contributions}
Conceptualization: I.L.
Formal analysis: I.L. (lead), all (supporting).
Methodology: I.L.
Investigation: I.L (equal), K.T. (equal)
Software: K.T.
Visualization: K.T.(lead), I.L. (supporting).
Data curation: K.T.
Supervision: I.L. (lead), H.N. (supporting).
Writing--original draft: I.L.
Writing--review \& editing: all.
Resources: H.N. (lead), all (supporting).
Funding acquisition: H.N. (lead), all (supporting).
The contribution of I.L. and K.T. to this work are equivalent.
	
	\section*{Declaration of generative AI and AI-assisted technologies in the manuscript}
	
    During the preparation of this work, the authors used Codex in order to improve the efficiency of the codes used for the numerical simulation. After using this tool, the authors reviewed and edited the content as needed and take full responsibility for the content of the published article.

    
	 \appendix
\renewcommand{\thesection}{\Alph{section}}

\numberwithin{equation}{section}
\counterwithin{figure}{section}

	\section{Stability analysis in the global coupling limit: deterministic case}\label{app.detkur}
	
	In this Appendix, we analyze the stability of the asymptotic dynamics of the higher-order Vicsek model in the absence of noise, i.e., Eq. (10) of the Main Text with $\eta=0$, in the global coupling limit, i.e., $r_0\to \infty$. In this case, the position of the particles are no longer relevant, and the dynamics are completely determined by the evolution of the alignment direction:
	\begin{eqnarray}
		\dot{\theta}_j&=&f(\theta_j)=\frac{\alpha }{N}\sum_{k=1}^N \sin(\theta_k-\theta_j)+ \frac{\beta}{N^2}\sum_{k,l=1}^N \sin(2\theta_k-\theta_l-\theta_j)\nonumber\\
		&=& \alpha r \sin(\psi-\theta_j)+ \beta r q \sin(\phi-\psi-\theta_j), \label{eq.kur}
	\end{eqnarray}
	where we keep the same symbols $re^{i\psi}$ and $qe^{i\phi}$ even if they are now computed summing over all particles. This equation is the Kuramoto model~\cite{Kuramoto_book,acebron2005kuramoto} with higher-order interactions~\cite{leon2025theory,leon2026symmetry}. In what follows, we analyze the stability of the states displayed by the system in the continuum, $N\to \infty$, limit, by employing common techniques from phase oscillator models~\cite{Kuramoto_book,acebron2005kuramoto,pikovsky2001synchronization,strogatz2000kuramoto,HMM93,KK2001,leon19,Buendia25}. 
		
	\subsection{Fully ordered phase}
	
	In the fully ordered phase, all particles are aligned with identical direction $\theta$, which we fix to zero without loss of generality, so that $re^{i\psi}=qe^{i\phi}=1$.  We can study the evolution of the perturbation of one particle, $\delta\theta$. Such perturbation does not affect the order parameters $re^{i\psi}$ and $qe^{i\phi}$, since, in the $N\to\infty$ limit, the perturbation of one particle does not affect the mean field. Hence, the perturbation evolves as
	\begin{equation}
		\dot{\delta\theta}=-(\alpha+\beta) \delta \theta_j.
	\end{equation}
	From the above equation, we can conclude that, if $\alpha+\beta\ge 0$, the fully-ordered is stable, as stated in the Main Text.
	
	\subsection{Bidirectional ordered phase}
	
	The bidirectional order is characterized by some of the particles having direction $\theta_A=0$, that we have fixed to zero without loss of generality, and the others moving in direction $\theta_B=\theta_A+\Delta$, with $\Delta=\pi$. This state corresponds to the two-cluster state observed in phase oscillator models~\cite{HMM93,KK2001,leon19}. We remark that the Kuramoto model with higher-order interactions also exhibits two-cluster states~\cite{tanaka2011multistable,leon2024}, which are fragile in the presence of noise~\cite{MK25}. In what follows, we borrow the techniques from those cited references to analyze the two-cluster state. Since we directly apply those methods, we invite the interested reader to consult the above references for more details and justification.
    
    %Ric: old text below:
    %%%This state corresponds to the two-cluster state observed in phase oscillator models~\cite{HMM93,KK2001,leon19} In this Supplementary Material, we apply those techniques without entering into the details.
	
	There are two relevant quantities that characterize the two-cluster state: the angle difference between clusters $\Delta$, and the proportion of particles $p$ with angle $\theta_A$. To study the stability of the two-cluster state, it is enough to consider only three perturbations: a perturbation of one particle in cluster $A$, $\delta \theta_A$, a perturbation of a particle in cluster $B$, $\delta \theta_B$, or a perturbation of the angle difference $\delta \Delta$. Each perturbation grows/decays exponentially according to its corresponding exponent:
	\begin{eqnarray}
		\lambda_A&=&-(-1+2p)(\alpha+\beta)\\
		\lambda_B&=&(-1+2p)(\alpha+\beta)\\
		\lambda_\Delta&=&\alpha-\beta+8\beta\left(p-\frac{1}{2}\right)^2.
	\end{eqnarray}
	
	Note that the stability of the two-cluster state depends on the proportion $p$ of oscillators in cluster $A$. It can be shown that the two cluster state is always unstable, since for all parameters there is at least one positive exponent, except for $p=1/2$ and $\alpha-\beta\ge0$, where it is neutrally stable. This seems to imply that bidirectional order is never stable. However, let us recall that this result is only true for global interactions, while bidirectional order appears when local interactions are considered. With a small modification to the argument, one can understand why it is stable for local interactions.
	
	If the interactions are local, $p$ is the average number of particles in direction $\theta_A$ interacting with a the perturbed particle. Because particles in direction $\theta_A$ move in parallel, they are constantly interacting with the same particles, while only occasionally with particles in the other direction. Meanwhile, particles moving in the direction $\theta_B=\theta_A+\pi$ interact more often with particles in that direction, rather than in the opposite direction $\theta_A$. This means that particles in direction $\theta_A$ see an effective $p>\frac{1}{2}$, while particles in direction $\theta_B$ see an effective $p<\frac{1}{2}$. If there are approximately the same number of particles in both directions, then for particles moving in direction $\theta_A$, $p=\frac{1}{2}+\delta p$ with small $\delta p$, while for particles in direction $\theta_B=\theta_A+\pi$, $p=\frac{1}{2}-\delta p$. Thus, the stability is given by
	\begin{eqnarray}
		\lambda_{A/B}&=&-\delta p(\alpha+\beta),\\
		\lambda_\Delta&=&\alpha-\beta+O(\delta p^2),
	\end{eqnarray}
	which are the exponents stated in the Main Text.

	\subsection{Disordered phase}
	For the analysis of the disordered phase, it is convenient to define the density of
	particles with direction $\theta$ between $\theta$ and $\theta+d\theta$ at time $t$, $\rho(\theta,t)$. Such a density evolves according to the continuity equation:
	\begin{equation}\label{eq.continuity}
		\partial_t \rho=-\partial_\theta(f(\theta)\rho),
	\end{equation}
	where in $f(\theta)$, c.f. Eq.~\eqref{eq.kur}, the order parameters $re^{i\psi}$ and $qe^{i\phi}$ are expressed in terms of the density:
	\begin{eqnarray}
		re^{i\psi}&=&\int_0^{2\pi} \rho(\theta,t) e^{i\theta}d\theta,\\
		qe^{i\phi}&=&\int_0^{2\pi} \rho(\theta,t) e^{i2\theta}d\theta.
	\end{eqnarray}
	We can then decompose the solution $\rho$ in its Fourier modes
	\begin{equation}
		\rho(\theta,t)=\frac{1}{2\pi}\sum_{k=-\infty}^\infty\rho_k(t) e^{ik\theta},
	\end{equation}
	with $\rho_{-k}=\rho_k^*$, to ensure that the density is real. Plugging the above decomposition into the continuity equation, each mode evolves as
	\begin{equation}\label{eq.contmodes}
		\frac{2}{k}\dot{\rho}_k=\alpha\left(\rho_{k-1}\rho_1-\rho_{k+1}\rho_{-1}\right)+\beta\left(\rho_{k-1}\rho_2\rho_{-1}-\rho_{k+1}\rho_{-2}\rho_1\right).
	\end{equation}
	
	The homogeneous disordered state corresponds to $\rho(t,\theta)=\frac{1}{2\pi}$, or, in Fourier modes, $\rho_0=1$, $\rho_{k\not=0}=0$. The evolution of the perturbation of such state $\rho_k=\delta \rho_k$ is governed by the deviation of the first mode:
	\begin{eqnarray}
		\dot{\delta\rho_1}=\frac{\alpha}{2} \delta\rho_1.
	\end{eqnarray}
	All other perturbations $\delta\rho_k$ grow or decay depending on the behavior of $\delta\rho_1$. Thus, the homogeneous disordered phase is stable when $\alpha<0$, as stated in the Main Text.
	
	Similarly, we can analyze the stability of the nematic disordered state. It is characterized by $\rho_0=1$, $\rho_{1}=0$, and all other modes can take any value. The evolution of the perturbation is solely governed by the perturbation of the first mode 
	\begin{eqnarray}
		\frac{d}{dt}\begin{pmatrix}
			\delta \rho_1\\
			\delta \rho_{-1}
		\end{pmatrix}
		=\frac{1}{2}
		\begin{pmatrix}
			\alpha-\beta |\rho_2|^2 & \rho_2(-\alpha+\beta)\\
			\rho_{-2}(-\alpha+\beta) & \alpha-\beta |\rho_2|^2
		\end{pmatrix}
		\begin{pmatrix}
			\delta \rho_1\\
			\delta \rho_{-1}
		\end{pmatrix}
		%    \dot{\delta\rho_1}=(\frac{\alpha-\beta |\rho_2|^2}{2} \delta \rho_1+(\frac{\alpha-\beta |\rho_2|^2}{2} \delta \rho_1
	\end{eqnarray}
that depends on the value of the nematic order parameter $\rho_{2}=qe^{i\phi}$. The eigenvalues of the matrix are:
	\begin{equation}
		\lambda=\alpha-q^2\beta\pm q |\alpha-\beta|,
	\end{equation}
    which are negative when
    \begin{equation}
    	\alpha\leq -q|\beta|.
    \end{equation}
	Notice that, since $q\in[0,1]$, this region lies inside the region where homogeneous disorder is stable, i.e., $\alpha<0$.
	
	\subsection{Partially ordered phase}
	The partially ordered solution in the deterministic case $\eta=0$ is the solution to the continuity equation Eq.~\eqref{eq.continuity} or \eqref{eq.contmodes} with $\rho_1\not=0$. If we go to a frame of reference such that $\phi=0$ or equivalently $\rho_2=q\in\mathbb{R}$, Eq.~\eqref{eq.contmodes} can be written as:
	\begin{equation}
		\frac{2}{k}\dot{\rho}_k=\left(\rho_{k-1}\rho_1-\rho_{k+1}\rho_{-1}\right)(\alpha+\beta\rho_2),
	\end{equation}
	with solution $\rho_2=q=-\frac{\alpha}{\beta}$, and any value for the other modes. Because $q\in[0,1]$, the solution only exists if $\alpha$ and $\beta$ have opposite signs and $|\alpha|\le|\beta|$.
	
	The stability of the solution is governed by the evolution of the perturbation of the second mode, that evolves according to:
		\begin{eqnarray}
		\frac{d}{dt}\begin{pmatrix}
			\delta \rho_2\\
			\delta \rho_{-2}
		\end{pmatrix}
		=
		\begin{pmatrix}
			\beta |\rho_1|^2 & -\beta \rho_1\rho_3\\
			-\beta \rho_{-1}\rho_{-3} & \beta |\rho_1|^2 &
		\end{pmatrix}
		\begin{pmatrix}
			\delta \rho_2\\
			\delta \rho_{-2}
		\end{pmatrix}.
	\end{eqnarray}
	Since the eigenvalues of the Jacobian are:
	\begin{equation}
		\lambda=\beta |\rho_1|(|\rho_1|\pm|\rho_3|),
	\end{equation}
	this phase is stable if $\beta < 0$.
	
	Merging all the previous restrictions, we obtain that the partially ordered phase is stable when 	$\beta<0$, $\alpha>0$ and $\alpha\leq-\beta$.
	
	\section{Stability analysis in the global coupling limit: stochastic case}\label{app.stochkur}
	
	In this Appendix, we analyze the stability of the asymptotic dynamics of the higher-order Vicsek model, Eq. (10) of the Main Text, in the global coupling limit, i.e., $r_0\to \infty$. This analysis provides the bifurcation diagrams Fig.~\ref{fig.bifdiagglobal} in the global case, capturing qualitative features of the bifurcation diagram of the Vicsek model with higher-order interactions, Fig.~\ref{fig.bifdiag}.
    
    In the global coupling limit, the position of the particles are no longer relevant, so the dynamics are completely determined by the evolution of the alignment direction:
	\begin{equation}
		\dot{\theta}_j=f(\theta_j)+\eta\xi_j(t).
	\end{equation}
	This equation is the stochastic Kuramoto model~\cite{Kuramoto_book,acebron2005kuramoto} with higher-order interactions. In the large system limit $N\to\infty$ limit, we can define a density of particles with angle $\theta$ that follows the Fokker-Planck equation
	
	\begin{equation}\label{eq.fokkerplack}
		\partial_t \rho=-\partial_\theta(f(\theta)\rho)+\frac{\eta^2}{2}\partial^2_\theta \rho,
	\end{equation}
	which, expressed in Fourier modes, becomes 
	\begin{equation}
		\frac{2}{k}\dot{\rho}_k=\alpha\left(\rho_{k-1}\rho_1-\rho_{k+1}\rho_{-1}\right)+\beta\left(\rho_{k-1}\rho_2\rho_{-1}-\rho_{k+1}\rho_{-2}\rho_1\right)-k \eta^2 \rho_k.
	\end{equation}
	
		\subsection{Disordered phase}
	We can analyze the stability of the disordered phase analogously to the deterministic case. The only disordered solution to the equation is $\rho(t,\theta)=\frac{1}{2\pi}$, or, in Fourier modes, $\rho_0=1$, $\rho_{k\not=0}=0$. The evolution of the perturbation of such state $\rho_k=\delta \rho_k$ is governed by the deviation of the first mode:
	\begin{eqnarray}
		\dot{\delta\rho_1}=\frac{\alpha-\eta^2}{2} \delta\rho_1.
	\end{eqnarray}
	Thus, the homogeneous disordered phase is stable when $\alpha<\eta^2$, as stated in the Main Text. 
	
	\subsection{Partially ordered phase}
	
	In order to analyze the partially ordered phase, we find the stationary solution of the Fokker-Planck equation Eq.~\eqref{eq.fokkerplack}. Since we are looking for a stationary solution, without loss of generality we can move to a frame of reference in which $\psi=0$. From Eq.~\eqref{eq.fokkerplack} we see that the solution verifies:
	\begin{equation}
			0=-\partial_\theta\bigg[\bigg(-\alpha r\sin\theta+\beta r q \sin(\phi-\theta)\bigg)\rho-\frac{\eta^2}{2}\partial_\theta \rho\bigg].
	\end{equation}	
	By making the expression inside the bracket parenthesis equal to zero, one can solve for $\rho$:
		\begin{equation}
		\rho(\theta)=\frac{1}{C} \operatorname{Exp}\bigg[\frac{2}{\eta^2} \big( \alpha r\cos\theta+\beta r q \cos(\theta-\phi)\big)\bigg],
	\end{equation}
	where $C$ is a constant to ensure $\rho$ is normalized. The normalization constant and other integrals below can be written in terms of the modified Bessel functions of the first kind:
	\begin{equation}
		I_n(x)=\frac{1}{2\pi}\int_{0}^{2\pi}e^{x\cos\theta}\cos(n\theta)d\theta,
	\end{equation}
	for integer $n$ values. After some calculations, the normalization constant is obtained  
	\begin{equation}
		C=2\pi I_0\left(|\gamma|\right),
	\end{equation}
	with $\gamma=\frac{2r}{\eta^2}(\alpha+\beta q e^{i\phi})$.
	
	Though we have formally the solution, we still need to know the values of $r$, $q$ and $\phi$. Those values can be obtained from the self-consistent conditions
	\begin{eqnarray}
		r&=&\int_{0}^{2\pi}\rho(\theta)e^{i\theta}d\theta,\label{eq.selfcondr}\\
		qe^{i\phi}&=&\int_{0}^{2\pi}\rho(\theta)e^{i2\theta}d\theta.\label{eq.selfcondq}
	\end{eqnarray}

	The integrals can be written again in terms of the modified Bessel functions of the first kind. The condition Eq.~\eqref{eq.selfcondr} yields:
	\begin{equation}
	r=\frac{I_1\left(|\gamma|\right)}{I_0\left(|\gamma|\right)}\frac{\gamma}{|\gamma|}.\label{eq.finrecrel}
\end{equation}
 	Because $r$ is real and positive, $\gamma$ has to be real and positive, so $\phi=0$ and $\gamma=\frac{2r}{\eta^2}(\alpha+\beta q)$. This condition simplifies all previous expressions. We can compute the integral in  the self-consistent condition, Eq.~\eqref{eq.selfcondq}, as:
 	\begin{equation}
 		q=\frac{I_2\left(|\gamma|\right)}{I_0\left(|\gamma|\right)}.
 	\end{equation}
 	The above expression can be further simplified by using the Bessel functions recurrence relation $I_n(x)-I_{n+2}(x)=\frac{2(n+1)}{x}I_{n+1}(x)$ and Eq.~\eqref{eq.finrecrel}:
 	\begin{equation}
 		q=1-\frac{\eta^2}{\alpha +\beta  q}.
 	\end{equation}
 	
 	There are only two solutions to this self-consistent condition:
 		\begin{equation} \label{eq.solqself}
 		q=\frac{-\alpha +\beta \pm\sqrt{\alpha ^2+2 \alpha  \beta +\beta ^2-4 \beta  \eta ^2}}{2 \beta },
 	\end{equation}
 	corresponding to a stable and unstable ordered state, respectively. Provided the value of $q$ and $\phi$, Eq.~\eqref{eq.selfcondr} can be solved to obtain $r$ and, thus, finding the partial ordered solution.
 	
 	This solution helps us to understand how and when the partially ordered phase emerges. From Eq.~\eqref{eq.solqself}, we observe a pair of solutions emerging from a saddle-node bifurcation when the discriminant is zero, i.e., $\alpha ^2+2 \alpha  \beta +\beta ^2-4 \beta  \eta ^2=0$, or, equivalently,
 	\begin{equation}\label{eq.eta_c}
 		\eta=\frac{|\alpha+\beta|}{2\sqrt{\beta}}
 	\end{equation}
 	corresponding to the abrupt order-disorder transition. Because the solution has to be positive, the saddle-node bifurcation has only physical meaning where $q>0$, corresponding to $\frac{-\alpha+\beta}{2\beta}>0$, or, equivalently, $\alpha<\beta$. 
 	
\begin{figure*}[t]
		\centering
		\includegraphics[width=2\columnwidth]{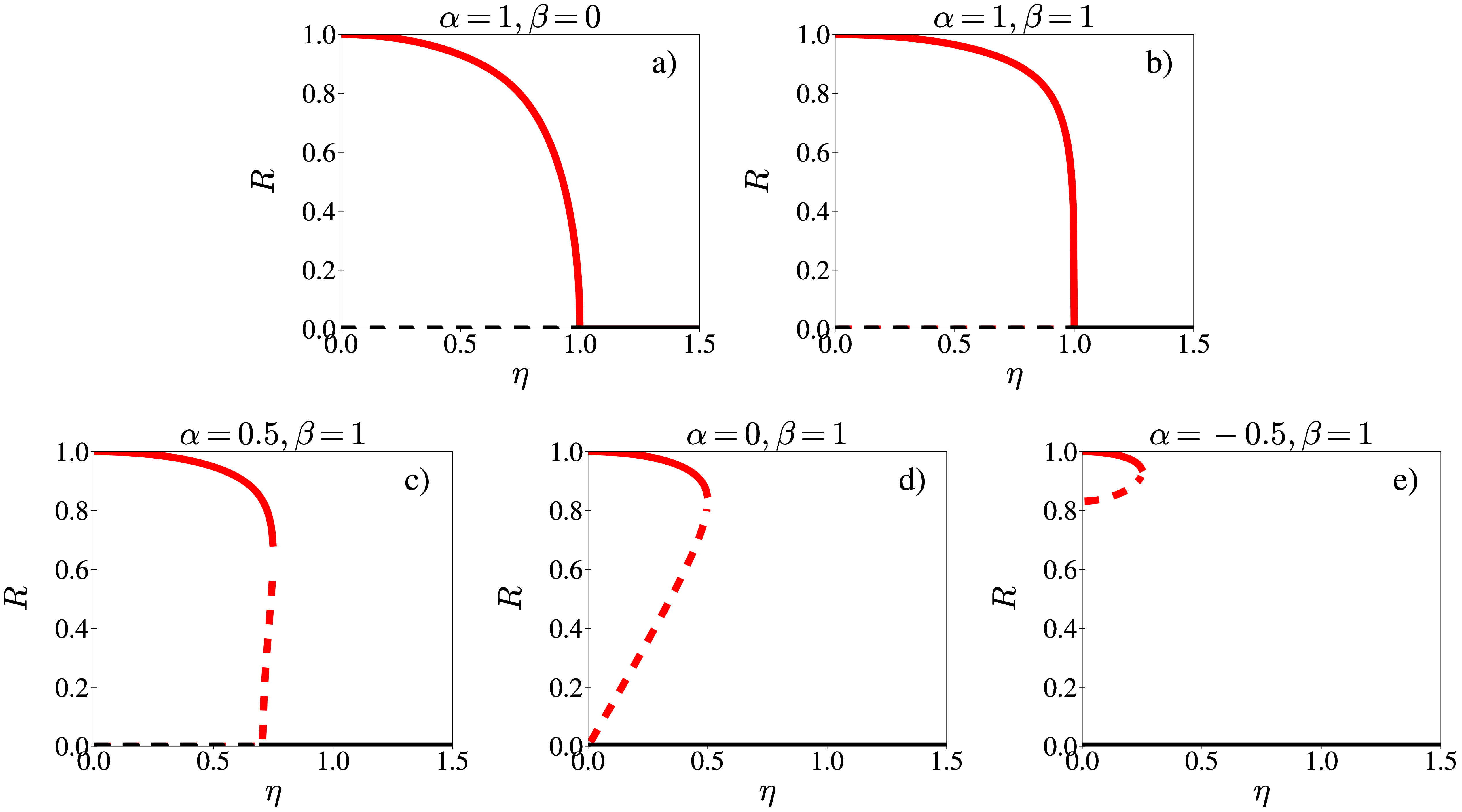}\\
		\caption{Global polar order parameter $R$ in the global coupling limit for different pairs $(\alpha,\beta)$ versus the noise intensity $\eta$. In panels (a), (b), the partially ordered state emerges when disordered state becomes unstable while in panel (c), (d) and (e) two partially ordered states collide in a saddle node bifurcation.}
		\label{fig.bifdiagglobal}
	\end{figure*}

 	From this study, we get to the conclusion that the partial ordered phase departs continuously from the disordered solution when the latter becomes unstable, i.e., $\eta=\sqrt{\alpha}$ for $\alpha\ge\beta$, Fig.~\ref{fig.bifdiagglobal} (a) and (b). Alternatively, a pair of stable and unstable partial order solutions emerge via a saddle-node bifurcation (abrupt order-disorder transition) when $\alpha<\beta$, Fig.~\ref{fig.bifdiagglobal} (c), (d), and (e). These conditions correspond to the stability boundary Eq.~\eqref{eq.partordbound} reported in the Main Text.

    Notice that the bifurcation diagrams of Fig.~\ref{fig.bifdiagglobal} for the global coupling limit are qualitatively similar to the ones of the higher-order Vicsek model, Fig.~\ref{fig.bifdiag}, where clear differences are due to the roughness of the global or mean-field approximation.  
 	
\end{document}